# SRS BUILDER 1.0: *An Upper Type CASE Tool for Requirement Specification.*

**Ardhendu Mandal**
*Department Of Computer Science and Application, University of North Bengal (N.B.U.), INDIA*
*am.csa.nbu@gmail.com*

**ABSTRACT**
*Software (SW) development is a labor intensive activity. Modern software projects generally have to deal with producing and managing large and complex software products. Developing such software has become an extremely challenging job not only because of inherent complexity, but also mainly for economic constraints unlike time, quality, maintainability concerns. Hence, developing modern software within the budget still remains as one of the main software crisis. The most significant way to reduce the software development cost is to use the Computer-Aided Software Engineering (CASE) tools over the entire Software Development Life Cycle (SDLC) process as substitute to expensive human labor cost. We think that automation of software development methods is a valuable support for the software engineers in coping with this complexity and for improving quality too. This paper demonstrates the newly developed CASE tools name "SRS Builder 1.0" for software requirement specification developed at our university laboratory, University of North Bengal, India. This paper discusses our new developed product with its functionalities and usages. We believe the tool has the potential to play an important role in the software development process.*

**KEYWORDS:** CASE tool, software development, SDLC, SRS, SE, FHD.

## 1. INTRODUCTION

Although, hardware costs are decreasing drastically, still the computers are not used extensively by the business organization because of the huge software cost. In recent years, Computer-Aided Software Engineering tools have emerged as one of the most important innovations in software development to manage the complexity of software development projects reducing its product cost. Using CASE tools over their SDLC process may reduce the developing cost significantly.

Although, almost all develop countries do use certain CASE tools, but its extensive use is still a dream because of their product cost. Although, big software giants do use their own developed or commercially available CASE tools in development software, but their quality, applicability, cost, availability remains as big question. The huge cost of such commercially available CASE tools made these outreach of the small software development companies. In this study, we are going to demonstrate newly developed requirement specification tool name SRS BUILDER version 1.0.

The rest of the paper is organized as follows: We started by defining software engineering, Computer Aided Software Engineering and CASE tools. Then, we have laid down the different types of CASE tools and advantage of using CASE tools with its limitation. In the later section we have discussed about SRS BUILDER 1.0 with its sample design.

## 2. COMPUTER AIDED SOFTWARE ENGINEERING (CSAE) and CASE TOOLS

In the following we are going to define some terminologies used in software engineering.

**A. Software Engineering (SE):** Before defining CASE, defining software engineering is justifiable. Although, the age of *Software Engineering* is quite old, but prior to its born, people do used to develop software. But because of some problems (discussion of these problems are beyond the scope of paper) faced later with those systems, software engineering emerged as a new subject.
The IEEE defines *software engineering* as [6]:
  i. The application of a systematic, disciplined, quantifiable approach to the development, operation, and maintenance of software; that is, the application of engineering to software.
  ii. The study of approaches as in (i).
More significantly, software engineering may be defined as the systematic approach to develop **quality** software **within both budget** and **time** constraints.

**B. Computer-Aided Software Engineering (CASE):**
CASE is an acronym that stands for **C**omputer-**A**ided **S**oftware **E**ngineering. Roughly, this is all about using computers at different phases of the SDLC process during the development and maintenance of software to assistance the development team. CASE provides the software engineer with the ability to automate manual activities and to improve engineering associated to software development. Basically, *it is all about using software to develop software.*

C. **CASE tools:** CASE tools are the tools that permit collaborative software development and maintenance. These tools are concerned with automated tools that aid in the definition and implementation of software systems.
Formal definition of CASE:
  i. "Individual tools to aid the software developer or project manager during one or more phases of software development (or maintenance)."
  ii. "A combination of software tools and structured development methodologies".

**D. Types of CASE tools:** Depending on the activities performed, CASE tools are primarily divided in to three categories. They are as follows:
i) **Upper CASE tools:** Primarily focuses on the System analysis and design phase of the SDLC.
ii) **Lower CASE tools:** Focuses on system implementation phase of SDLC.
iii) **Integrated CASE tools:** It helps in providing linkages between the lower and Upper CASE tools.

## 3. ADVANTAGES OF USING CASE TOOLS

With the growing importance of CASE tools, more steps in the SDLC are being automated. However, a complete automated software facility for all steps is still a dream. Presently available CASE tools cover only a certain modules of the general CASE facility.

The benefits of using CASE tools are as follows:
a. Increasing Productivity.
b. Product Quality improvement.
c. Development Cost Reduction.
d. Effort Reduction: Different studies carried out to measure
   the impact of CASE put the effort reduction about 30-40% [4].
e. Reduction in Development Time.
f. Reduce the drudgery and working style in a software engineer's work.
g. *Create good quality documentation. Our, proposed tool basically focuses on this particular aspect of computer aided software engineering.*
h. Create maintainable system.
i. Providing a uniform platform for software/system developers to present information and knowledge compactly for ease of communication (Banker & Kauffman, 1991; Church & Matthews, 1995; Orlikowski, 1989).

## 4. USAGE OF CASE TOOLS: A RESEARCH REPORT

CASE (Computer-Aided Software Engineering) tools are supposed to increase productivity, improve the software product quality and make Information Systems development a more enjoyable task [2]. However, they have been failing to deliver the benefits they promise [1]. The results of the research carried out by Kemerer, C. F. [3] regarding the usage of CASE tools after introduction are shown in the *Figure 1*.

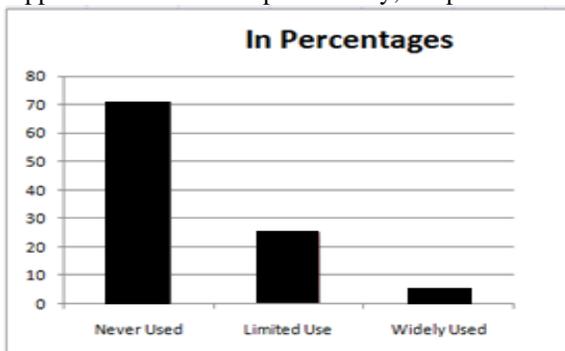

Figure 1: Percentage of Usage of CASE tools after introduction

This results about CASE tool usage raise a significant question, "Why the percentage of widely used case tools are too low (5% only)?" The answer to the question is the limitations with the CASE tools discussed in the next section.

## 5. LIMITATION OF CASE TOOLS

CASE tools are still in limited use because of the following limitations:
a) CASE tools don't support automatic development of functionally relevant system.
b) It force system analyst to follow a prescribed methodology.
c) It may change the system analysis and design process.
d) Poorly supported expected functionality from them.
e) Huge cost of the CASE tools.
f) Lack of Concern in CASE tool usage.
g) Bad quality of the CASE tools.
h) CASE tools are complex because they offer a large array of options and support for software development activities.
i) Cheap Labour cost: In developing countries like India and other underdeveloped countries, cheaply available human resources remain as one of the biggest reasons for not using the costly CASE tools.

## 6. MOTIVATION FOR UNDERTAKING THE RESEARCH PROJECT

In addition to usage in industry by practitioners, CASE tools are employed by educators to teach students software development skills. Evaluating and selecting a CASE tool for a specific systems development course is quite difficult.

While teaching the paper software engineering at university (NBU), it was found that the students are very much confused about CASE tools. The significant reasons for the same are basically widely available complex poor quality CASE tools which are quite expensive to purchase although. Some modules are good in some if not in all. The students do not feel interesting to use them. Then we under took the project to develop a new customised CASE tool for requirement specification supporting IEEE specification as per the students need that will help the student to improve their skill and have a clear vision about it. We start with the development of CASE tool to specify the system requirements to generate the System Requirement Specification (SRS) document. The outcome of our attempt is the so named, "***SRS BUILDER 1.0***"- the CASE tools to generate the SRS document. Moreover, we are planning to distribute the newly developed CASE tool to the educational institutions on demand almost free of cost with a nominal transportation cost for the sake of student community.

## 7. SRS BUILDER 1.0: THE NEW REQUIREMENT SPECIFICATION TOOL

As from the SLC models, we know that after the requirements are gathered by the system analyst, the analysed and finalised requirements need to be specify in the SRS document. The SRS document is then provided to the software development team for next phase of the development.

a) ***SDLC Model Followed:*** We have followed the **BRIDGE** software development process model [5] to develop the *SRS BUILDER 1.0*.

b) ***Product Quality Features:***
→ Support to IEEE specification for SRS writing format along with the flexible other customized specifications as per your organizational need.
→Good Graphical User Interface
→Easy to use
→Easy Installation
→ Incorporated User Documentation
→Easily Available
→Compatible
→Minimal System Requirements for Installation
→Validated

c) ***System Specification:***
**Front End:** Visual Basic 6.0
**Back End:** MySQL
**Operating System:** Windows XP, Windows Vista.
**Memory Requirement:** 15 KB

## 8. FUNCTION HIERARCHY DIAGRAM (FHD) OF SRS BUILDER 1.0

The Function Hierarchy Diagram (FHD) of the SRS BUILDER 1.0 is shown below in *Figure 2*.

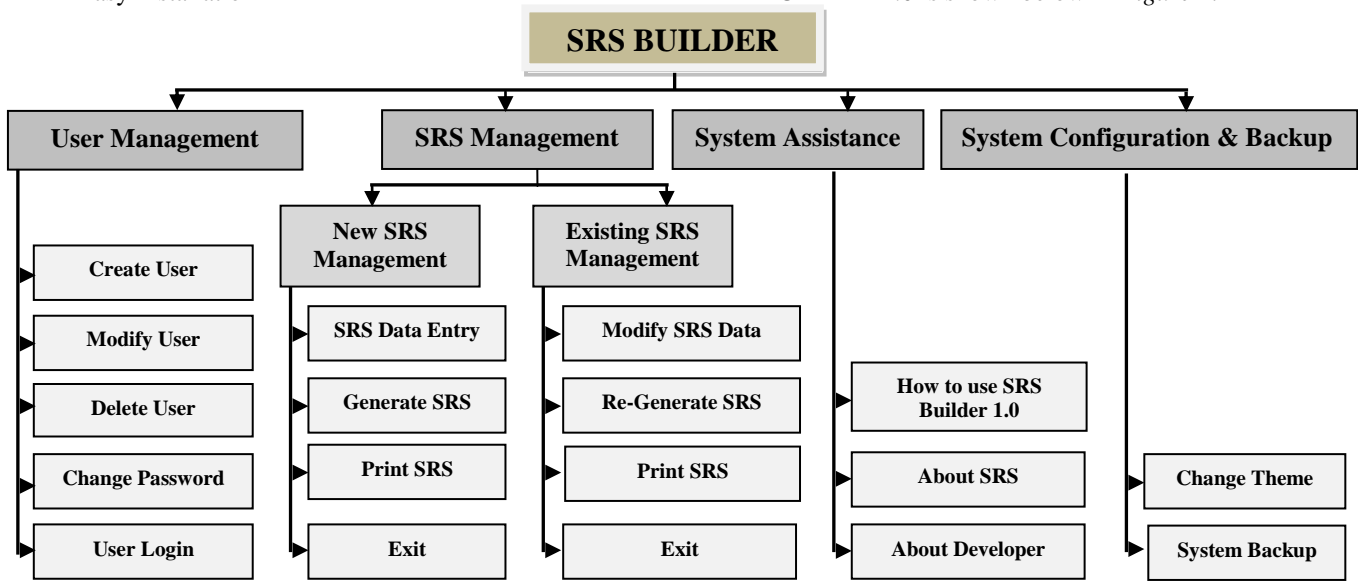

**Figure 2:** *Function Hierarchy Diagram (**FHD**) of SRS Builder 1.0*

## 9. SAMPLE SRS ORGANIZATION GENERATED BY SRS BUILDER 1.0

A typical SRS generated by the SRS BUILDER 1.0 for the ATM system is shown below in *Figure 3*. This is not a complete and correct SRS for the intended system, but a sample used only to show the SRS organization generated by the tool. The font size and spacing has been changed to accommodate in the paper keeping the context organization unchanged.

**Project Title: SRS ATM**

**Project Id: 1**

<div align="center">

**SOFTWARE REQUIREMENT SPECIFICATION**

**<u>Introduction</u>**

</div>

**Purpose:**

This document describes the software require……

**Scope:**

The software supports a computerized banking …….

**Definition:**

• Account:

A single account at a bank against which transaction…………..

**Intended Audience:**

The intended audience of this SRS consists of:

• Software designers….

**Reference: NA**

**Overview: NA**

**Document Conventions: NA**

## Overall Description

**Product Perspective:**

An automated teller machine (ATM) is a……. customer is identified by inserting a plastic ATM card …………………

**Product Function:**

1. Get Balance Information…..

**User Characteristics:**

Open to all authorized users…….. Customers are simply members of the public with no special training……

**Operating Environment:** Ability to read the ATM card………………

**General Constarints: NA**

**User Documentation: NA**

**Assumptions Dependencies:** Hardware never fails ………….

## Specific Requirements

N.A…………………

## External Interface Requirements

**User Interface:**

The customer user interface should be intuitive, such that 99.9% of all new ATM users are able……..

**Hardware Interface:**

• Ability to read the ATM card....................

**Software Interface:**

• State Bank…………

**Communication Interface:**

• List of Communicational interface equirements …………

**Functional Requirements:**

• List of functional requirements …………

**Behavioural Requirements:**

• List of behavioural requirements …………

<u>**Other Non-functional Requirements**</u>

**Performance Requirements:**

• It must be able to perform in adverse conditions like high/low temperature etc. …………

**Safety Requirements:**

• Must be safe kept in physical aspects, say in a cabin ………….

**Security Requirements:**

• Users accessibility is censured in all the ways ……………….

**Software Quality: NA**

**Other Requirements: NA**

**SYSTEM REQUIREMENTS SPECIFICATION for ATM Withdrawal**

**Submitted by:**

______________________________________   ____________________
Program Manager/Functional Project Officer                Date

**Coordination:**

______________________________________   ______________________
Director, Applications Architecture                       Date

______________________________________   ______________________
Director, Engineering                                     Date

______________________________________   ______________________
Test Director                                             Date

**Approved by:**

______________________________________   ______________________
Functional Manager                                        Date

**Figure 3:** A typical SRS organization generated by SRS BUILDER 1.0

## 10. CONCLUSION

We may conclude the paper by pointing out that, this CASE tool will play an important role to the software developers and learners to use and understand the utility of the CASE tool in today's complex software projects. Also, as we mentioned earlier, interested educational institutions and organizations may contact the author for the CASE tool for their usage.

## 11. FUTURE WORK

SRS BUILDER 1.0 CASE tool has been tested and validated properly. In future, we propose to enhance the capabilities of the present version by appending the functionalities to design the various UML diagrams.